\documentstyle[12pt]{article}
\begin{document}
\def\strut{\rule[-.5cm]{0cm}{1cm}}
\def\dspace{\baselineskip = .30in}


\title{Zeroing In On the Top Quark, LSP and Scalar Higgs Masses
\thanks{Supported in part by Department of Energy
Grant No. DE-FG02-91ER40626.}}

\author{{\bf B. Ananthanarayan}\thanks{Present Address: Institut de
physique th\'{e}orique, Universit\'{e} de Lausanne, CH-1015,
Lausanne, Switzerland}\\
Theory Group,\\
Physical Research Laboratory\\
Navrangpura, Ahmedabad 380009, India
\\\\
{\bf Q. Shafi}\\
Bartol Research Institute\\
University of Delaware\\
Newark, DE 19716, USA}

\date{ }
\maketitle

\begin{abstract}
We estimate the top quark, lightest sparticle (LSP) and scalar higgs
masses within a supersymmetric grand unified framework in which
$\tan\beta \simeq m_t/m_b$ and the electroweak symmetry is
radiatively broken. The requirement that the calculated $b$ quark
mass lie close to its measured value, together with the cosmological
constraint $\Omega_{LSP} \approx 1$, fixes the top quark mass to be
$m_t(m_t) \approx 170 \pm 15\ GeV$. The LSP (of bino purity
$\stackrel{_>}{_\sim} 98\%)$ has mass $\sim 200 - 350\ GeV$. In the
scalar higgs sector the CP-odd scalar mass $m_A \stackrel{_<}{_\sim}
220\ GeV$. With $m_A \stackrel{_>}{_\sim} M_Z$, as suggested by the
decay $b \rightarrow s\gamma$, we find $M_Z \stackrel{_<}{_\sim}
m_{h^0} (m_{H^0}) \stackrel{_<}{_\sim} 140 (220)\ GeV$ and $120\ GeV
\stackrel{_<}{_\sim} m_{H^\pm} \stackrel{_<}{_\sim} 240\ GeV$.
\end{abstract}
\newpage

\dspace
The lower bounds on the masses of the top quark and the scalar higgs of the
standard model now stand at $113\ GeV$\cite{cdf} and $60\ GeV$\cite{aleph}
respectively. Estimates of the top quark mass based on a global
analyses of the electroweak data suggest the value $150^{+19+15}_{-24-20}\;
GeV$\cite{lang}. This approach does not, however, provide any precise
indication
of where the higgs mass may lie, and an answer almost certainly
requires an extension of the standard model. Presumably supersymmetry
is the way
to go, but the minimal supersymmetric standard model (MSSM)\cite{hn}
introduces two new parameters in the
scalar higgs sector, usually parameterized as $m_A$ (the tree level
mass of the CP-odd scalar boson) and $\tan\beta$ (the ratio of the
two vacuum expectation values (vevs), needed to provide masses to
up-type and down-type quarks). Knowing $m_A$ and $\tan\beta$ allows
one to estimate the tree level masses of the scalar higgs sector of
MSSM. An important challenge to the model builder therefore is to
find a framework in which something definite can be said about these
as well as a number of other undetermined parameters of MSSM.

It has been emphasized[5,6] that embedding the MSSM in
grand unified theories (GUTs) based on gauge groups such as $SO(10)$
or $SU(3)_c \times SU(3)_L \times SU(3)_R$ (but not $SU(5)$!) can
constrain $\tan\beta$ to a region close to $m_t/m_b \gg
1$. With $m_A \stackrel{_>}{_\sim} m_Z$, as suggested by the decay $b
\rightarrow s\gamma$, one readily understands in this scheme why no
higgs scalar has been found
at LEPI. The tree level mass of the lighter CP-even scalar mass is
approximately $M_Z$, and radiative corrections can increase it
further. In order to estimate the latter one needs a reliable
estimate of the top quark mass, a quantity closely related to the important
issue
of radiative electroweak breaking[7].

It was first pointed out in [5] that with $\tan\beta \approx
m_t/m_b$, the requirement that $m_b(m_b) = 4.25 \pm 0.10\ GeV$ also
helps constrain the top mass within a relatively narrow range. This
has subsequently been refined by several authors\cite{hrs}, and the results
can be summarized as follows. With $\alpha_s (M_Z) = 0.115 \pm 0.005$
and requiring that $m_b(m_b) \approx 4.25 \pm 0.10\ GeV$, the top mass
satisfies $155\ GeV \stackrel{_<}{_\sim} m_t(m_t) \stackrel{_<}{_\sim}
200\ GeV$. As we will see, the radiative breaking scenario, coupled
with the constraint $\Omega_{LSP}$, imposes a somewhat more stringent upper
bound on the top mass, to wit, $m_t(m_t) \stackrel{_<}{_\sim} 185\
GeV$. That is, we expect $m_t(m_t)$ to be in the $155-185\ GeV$ range.

In this paper we refine and extend the discussion of ref. [6]
to a point where several crucial mass
parameters, including the top quark, LSP and scalar higgs masses, can
be reliably predicted. An essential new
ingredient is the constraint that the LSP contribution to the energy
density of the universe be cosmologically significant, $\Omega_{LSP}
\approx 1$. The simplest scenario of large scale
structure formation (with a Harrison-Zeldovich spectrum of primordial
density fluctuations) to survive after COBE is based on the idea that the dark
matter is a
mixture of cold and hot components[9]. Roughly 65-70\% of the
matter is cold, 25-30\% is
hot, and the rest resides in the baryons.
We therefore require that the LSP makes a substantial, if not the sole,
contribution to the cold component. We find that consistency between
radiative electroweak breaking and the constraint $\Omega_{LSP}
\stackrel{_<}{_\sim} 1$, requires that the top quark mass $m_t(m_t)
\stackrel{_<}{_\sim} 185\ GeV$. It turns out that the LSP (with bino
purity $\stackrel{_>}{_\sim} 98\%)$ has mass in the range of $200 -
350\ GeV$. (For bino dark matter see ref. [10]; we assume a Hubble
constant of $40 - 50\ km\ s^{-1} Mpc^{-1}$.) Upper bounds on the
masses of the higgs bosons of MSSM are also derived. An essential
ingredient here is the constraint $m_A \stackrel{_<}{_\sim} 3\ M_W$ obtained
for the mass of the CP-odd scalar boson. This
enables us to put upper bounds on all of the scalar higgs masses.

The large $\tan\beta (\approx m_t/m_b)$ model of interest here can be
motivated, as indicated earlier, in a variety of grand unified frameworks
including
$SO(10)$ and $SU(3)_c \times SU(3)_L \times SU(3)_R$. In $SO(10)$, for
instance, the requirements that the third family charged fermions
acquire masses primarily from the coupling ${\bf 16} \times {\bf 16}
\times {\bf 10}$[5], and that the dominant components of the light Higgs
doublet pair come from the {\bf 10} plet, lead to the (asymptotic) relation
$\tan\beta = m_t/m_b$ valid above the GUT scale $M_X$. Below $M_X$,
$\tan\beta$ lies close to (actually just below) $m_t/m_b$.

The starting point of our computation is the estimation of the
unification scale through integration of the one-loop renormalization
group equations (RGE) for the gauge couplings.  Note that the susy
renormalization group equations are switched on above the scale
$Q_0 (\sim 1\ TeV)$, which is comparable to the
squark masses.  Knowing $M_X$ and the unified gauge
coupling $\alpha_G$, we proceed to scan the parameter space by varying
the boundary
values of the various quantities $M_{\frac{1}{2}}$ (universal gaugino
mass), $m_0$ (universal scalar mass), $A$ (universal
tri-linear scalar coupling) and $h$ (unified Yukawa
coupling).  The quantity $\tan\beta$ is computed at $Q_0$, using the
measured value $m_\tau(m_\tau) = 1.78\ GeV$.
Our analysis is based on the use of the renormalization group
improved tree level scalar potential whose relevant part takes the
standard form[4,7]
\pagebreak

\begin{eqnarray}
V_0(Q)=\mu^2_1\mid H^0_1\mid^2 + \mu^2_2 \mid H^0_2\mid^2 + \mu^2_3
(H^0_1 H^0_2 + h.c.)\nonumber\\
+ \frac{g^2_1 + g^2_2}{8} \left( \mid H^0_1 \mid^2 - \mid H^0_2
\mid^2 \right)^2
\end{eqnarray}

\noindent
Here the various parameters are evolved according to one loop
RGE from $M_X$ to $Q_0$. It has been pointed out in the literature that
a quantitative discussion
of the electroweak breaking requires the full one-loop effective
potential. However, a reasonable estimate is obtained with $V_0$ if
$Q_0$ is chosen to be comparable to the stop (sbottom) masses[11]
which translates to $Q_0 \approx 1 - 1.5 TeV$.

The spontaneous breaking of the electroweak gauge symmetry imposes
the constraint

\begin{equation}
\mu^2_1\mu^2_2 < \mu^4_3
\end{equation}

\noindent
Furthermore, the boundedness of the scalar potential yields

\begin{equation}
\mu^2_1 +\mu^2_2 \geq 2\mid\mu^2_3\mid
\end{equation}

\noindent
Another possible constraint at scale $Q_0$ is provided by the
requirement[12]

\begin{equation}
\label{komatsu}
m^2_{H_{2}} + m^2_{\tilde{L}} > 0
\end{equation}

\noindent
where $m^2_{H_{2,}H_{1}} = \mu^2_{2,1} - \mu^2 (\mu$ denotes the
supersymmetric higgs(ino) mass), and $m^2_{\tilde{L}}$ is the squared
mass of the (third family) SU(2)$_L$ slepton doublet.

Although this is derived within the framework of the renormalization
group improved tree level scalar potential, it has been argued in [11]
that eq. (4) is ineffective once account is taken of the fact that
eq. (1) is only trustable if the one loop radiative corrections are
sufficiently small. Consequently, in what follows, we will ignore the
constraint in eq. (4) [In contrast to our earlier work, ref. [6], in
which (4) was kept and we found $m_t(m_t) \stackrel{_<}{_\sim} 160\
GeV$].

It follows from the relation

\begin{equation}
\mu^2_1(Q_0) - \tan^2\beta\mu^2_2(Q_0) = \frac{1}{2} M^2_Z
(\tan^2\beta -1)
\end{equation}

\noindent
 that for $\tan^2\beta \gg 1$, $\mu^2_2$ must lie close to
its lower bound $-M^2_Z/2$.
If we know $m^2_{H_{1}}$ and $m^2_{H_{2}}$ at $Q_0$, we can
determine $\mu^2(Q_0)$ from the approximate relation

\begin{equation}
\mu^2(Q_0) \approx - m^2_{H_{2}} - \frac{M^2_Z}{2} +
\frac{m^2_{H_{1}}}{\tan^2\beta}
\end{equation}

\noindent
This can be used to estimate $\mu^2_3$:

\begin{equation}
\mu^2_3(Q_0) \approx \tan\beta [- m^2_{H_{2}} - \mu^2 -
\frac{M^2_Z}{2}]
\end{equation}

\noindent
Note that $B \equiv \mu^2_3/\mu$ is then also determined.

The appropriate one loop renormalization group equations for the
relevant quantities are well known[7] and will not be given
here. Following the conventions of the last reference in [7],
the signs of the Yukawa couplings $(h_t,\ h_b,\ h_\tau)$ and of $\mu$ are taken
to be  positive.
The mass spectrum could still display dependence on the signs of
$\tan \beta, \, M_{\frac{1}{2}}$ and $A$.  It follows from eqs. (5),
(6) and (7) that the quantity $B(Q_0)$ is sensitive to the sign of
$\tan\beta$.  However, the magnitude of $B(Q_0)$ always turns
out to be small enough, so that this dependence is not significant for
our purposes.  Similarly, even though the neutralino, chargino
and sfermion
mass matrices\cite{hn} display
some dependence on the sign of $\tan\beta$, it turns out to be not
very significant.  We will therefore assume that $\tan\beta > 0$.
Furthermore, our analysis shows that deviations of $A(M_X)$
from zero can never be large, so that
the bounds depend on $A(M_X)$
only in a minor way.  With $A(M_X) \approx 0$, the
sparticle spectrum is unchanged if $M_{\frac{1}{2}}\rightarrow
-M_{\frac{1}{2}}$. It is important to point out that even with $A(M_X)
\approx 0$, the quantities $A_t$ and $A_b$, when evaluated at
the scale $Q_0$, lie in the TeV range.

Let us now see how the bounds on $m_t(m_t)$
are obtained.  The first thing to note is that
for each $h$ (the unified Yukawa coupling at $M_X$), there is a lower
bound on $M_{\frac{1}{2}}$
for a satisfactory realization of the scenario.  We call this lower bound
$M_{\frac{1}{2}}^{min}|_{m_{A=0}}$, where $m_A^2 = \mu_1^2 + \mu^2_2$ is
the mass squared of the CP odd boson.
Once this is found, consider some fixed $M_{\frac{1}{2}}$
$> M_{\frac{1}{2}}^{min}|_{m_{A=0}}$, and find the value of $m_0$ above which
$m_A^2$ turns negative, keeping the other parameters fixed. As $m_0$ is
decreased below this value $m_A$ increases.  Thus, for fixed $M_{\frac{1}{2}}$
and $h$, there is
a maximum value, $m_A^{max}$, which can be obtained by lowering $m_0$.
By doing this a value of $m_0$ is encountered below which one of the
staus is the lightest sparticle, which is unacceptable.

For fixed $h$, for smaller values of $M_{\frac{1}{2}}$, the value of
$m_A^{max}$ is smaller.  It is possible, therefore, for fixed $h$
and a given $m_A$ to find  $M_{\frac{1}{2}}^{min}|_{m_{A}}$ (and a
corresponding $m_0^{min}|_{m_{A}}$).
Varying $h$ yields a profile for $M_{\frac
{1}{2}}^{min}|_{m_{A}}$ (and a corresponding one for
$m_{0}^{min}|_{m_{A}}$), which can be plotted against $m_t(m_t)$ (corresponding
to $h$). An example is shown in Fig. 1, corresponding to $m_A = m_Z$. In most
of the region of interest, with larger values
of $h$ and fixed $m_A$, the quantity $M_{\frac{1}{2}}^{min}|_{m_{A}}$
increases. We can see from the shape of this profile that with
increasing $m_t(m_t)$, to obtain larger $m_A$ requires increasing values of
$M_{\frac{1}{2}}$. In other words, if one plots, for given
$m_A$, contours of $M_{\frac{1}{2}}^{min}|_{m_{A}}$ (and a corresponding
$m_0^{min}|_{m_{A}})$ as functions of $m_t(m_t)$, then
for given $m_{A_{1}}, m_{A_{2}}$ with $m_{A_{1}} > m_{A_{2}}$,
contours corresponding to
$m_{A_{1}}$ lie above those corresponding to $m_{A_{2}}$. Note that
in this work we will take $\alpha_s (M_Z)$ in the range 0.11 to
0.12.

Next we discuss the role played by cosmological considerations
in restricting the value of $m_t(m_t)$.
As argued above, as $h$ increases the value of
$M_{\frac{1}{2}}^{min}|_{m_{A}}$
required for obtaining a satisfactory
scenario also rises.  It turns out that for
$m_t(m_t) \stackrel{_{>}}{_{\sim}} 145\ GeV$, the LSP essentially
consists of the
bino with a small $(\stackrel{_<}{_\sim} 2\%)$ higgsino
component (Fig. 2).
According to Ref. [10], the bino mass should be
below $350\ GeV$ in order that $\Omega_{LSP}$ does not exceed unity,
which means that the common gaugino mass at $M_X$ should
not exceed $\sim 800\ GeV$.  Using Fig. (1), this translates into an
upper bound $m_t(m_t) \stackrel{_{<}}{_{\sim}} 185\ GeV$ (for $\alpha_S (M_Z) =
0.12)$. [With $\alpha_S(M_Z) = 0.11, Q_0 = 1\ TeV$, the upper bound on
$m_t(m_t)$ is reduced by a few percent.]

In summary, with $\tan\beta \approx m_t/m_b$, the radiative
electroweak breaking scenario, combined with cosmological
considerations of the LSP, predict that the top quark is to be found
in the mass range $145\ GeV \stackrel{_<}{_\sim} m_t(m_t)
\stackrel{_<}{_\sim} 185\ GeV$. Furthermore, the LSP (with bino
purity $\stackrel{_>}{_\sim} 98\%)$ mass is estimated to be $\sim 200
- 350\ GeV$. Some comments about the scale-dependence of the results
are in order. By varying $\alpha_s (M_Z)$ between 0.11 and 0.125 and
$Q_0$ between 0.5 and $TeV$, we find that $M_{\frac{1}{2}}^{min}$ and
$m_0^{min}$ vary by $\sim 5 - 10\%$. This alters the upper bound on
$m_t(m_t)$ by about $\pm 2\ GeV$. (Similar considerations later apply
to the scalar higgs masses.)

As indicated earlier, with $\tan\beta \approx m_t/m_b$, an
independent estimate of $m_t(m_t)$ is obtained by requiring that the
calculated $m_b(m_b)$ is close to the value $4.25 \pm 0.10\ GeV$ given
in ref. [13]. The most recent estimates[8] yield $m_t(m_t) \approx
155 - 200\ GeV$.

We can now combine the two independent estimates of $m_t(m_t)$, from
considerations of $m_b(m_b)$, and from radiative electroweak breaking
coupled with the requirement that $\Omega_{LSP} \approx 1$, to yield
the overlap range of $155 - 185\ GeV$ for $m_t(m_t)$. Taking as
`central' values $\alpha_s (M_Z) = 0.115$ and $m_b(m_b) = 4.25\ GeV$, we
predict $m_t(m_t) = 170 \pm 15\ GeV$.

Let us now consider the important issue of the scalar higgs masses of
the MSSM. In the class of models under discussion, the parameter
$m_A$ is constrained by the approximate upper bound of $3\ M_W$,
corresponding to $\alpha_S (M_Z) = 0.12$. [With $\alpha_S (M_Z) = 0.11$
this is reduced by $\approx 10\%$.] Thus, we have
an unusually well specified version of the MSSM. Following
the first paper in Ref. [14],
we have estimated the radiative corrections to the scalar higgs
masses and some of the results are presented in Figs. 3, 4 and 5. In
Fig. 3 we show how $m_{h^\circ}$ varies with $m_A$ for typical values of
the parameters. With $m_t(m_t) \approx 170\ GeV$ say, the maximum
value is $m_{h^\circ}^{max} \approx 130\ GeV$. In Fig. 4 we
display the dependence of $m_{h^\circ}^{max}$ on $m_t(m_t)$ which
yields the absolute upper bound on the former of $\approx 140\ GeV$.
In Fig. 5 the dependence
on $m_t(m_t)$ of the scalar masses $m_{H^\circ},
m_A$ and $m_{H^\pm}$ are displayed. The results are $m_A^{max}
\approx m_{H^\circ}^{max} \stackrel{_<}{_\sim} 220\ GeV$ and
$m_{H^\pm}(\equiv \sqrt{m_A^{2} +m_W^2}) \stackrel{_<}{_\sim} 240\ GeV$
(for $\alpha_S (M_Z) = 0.12$). Somewhat lower values are obtained
with $\alpha_S (M_Z) = 0.11$.

In order to explain the shape of the curve determining
$m_{h^\circ}^{max}$ (Fig. 4), we first draw attention to the fact
that a sufficiently large value of $m_A$ is needed to ensure
that the lighter (CP-even) higgs $h^\circ$ receives a substantial part of
the radiative corrections to the tree-level relations in
the presence of large Yukawa couplings (see Fig. 3).
[Recall that the lightest higgs receives substantial
corrections only in the ``Weinberg-Salam'' limit, $m_A>>M_Z$.]
This explains, for `sufficiently small' $m_t(m_t)$, the near linear
rise in $m_{h^\circ}^{max}$ (Fig. 4).  However, for $m_t(m_t)
\stackrel{_>}{_\sim}
180\ GeV$, the parameter space rapidly shrinks and $m_A^{max}$
declines sharply (Fig. 5).  As a result, the radiative
corrections to the mass of the lightest higgs begin to diminish in
importance, and finally $m_{h^\circ}^{max}$ becomes nearly degenerate
with $m_A^{max}$.  This is seen in Fig. 4 in the
initial linear rise of $m^{max}_{h^\circ}$,  followed by
a slow down and culminating in a sharp fall. Meanwhile, as $m_t(m_t)$
increases, the maximum tree-level mass of the heavier CP-even neutral higgs
$H^\circ$ becomes nearly degenerate with $M_Z$ (whereas it is nearly degenerate
with
$m_A^{max}$ for the lower range of allowed $m_t(m_t)$) and receives the
larger share of the radiative corrections. Thus, $m_{H^\circ}^{max}$
reaches a minimum as $m_t(m_t)$ increases and then rises (Fig. 5).

In order to bound $m_{h^\circ}$ from below we note that a lower bound on $m_A$
more stringent than from the LEP data is
obtained from considerations of the decay $b \rightarrow s\gamma$.
The CLEO collaboration[15] has recently reported the first
observation of this process at a rate consistent with the standard
model estimates. For a top quark mass of $150\ GeV$, theoretical estimates
[16]
(prior to the discovery) suggested that $m_A \stackrel{_>}{_\sim}
M_Z$ (perhaps even $\stackrel{_>}{_\sim} 130\ GeV$!). Combining this
with the above considerations, we are led to an estimate of the $h^\circ$
mass in the  range $92 - 140\ GeV$. The model therefore ``explains'' why the
higgs
scalar $h^\circ$ has not been seen at LEPI. It also may elude an
upgraded LEPII unless $m_A$ happens to lie close to $M_Z$. For
instance, with $m_A \approx 100\ GeV$ and $m_t(m_t) = 170\ GeV,\
m_{h^0} \approx 100\ GeV$. (See Fig. 3).

To conclude, the MSSM raises many more questions than it answers. We
have argued that a supergravity/grand unified approach in which
$\tan\beta$ is close to $m_t/m_b$ provides an attractive and perhaps
even the most predictive extension. Indeed, we are able to shed light
on a number of the most fundamental parameters of MSSM. This includes
a rather precise determination of the top quark mass, $(155 - 185\ GeV)$,
a narrow allowed range for the LSP mass $(\sim 200 - 350\ GeV)$, and a
stringent upper bound $(\approx
220\ GeV)$ on $m_A$ which forces the scalar higgs to all have masses
below about $250\ GeV$. The sparticle mass spectrum is also quite
constrained. The lightest charged sparticles include a stau
$(m_{LSP} \stackrel{_<}{_\sim} m_{\tilde{\tau}_{1}} \stackrel{_<}{_\sim}
580\ GeV)$ and a chargino $(370\ GeV \stackrel{_<}{_\sim}
m_{\chi_{1}^\pm} \stackrel{_<}{_\sim} 670\ GeV)$. The heaviest states
include the gluino $(1 TeV \stackrel{_<}{_\sim} m_{\tilde{g}}
\stackrel{_<}{_\sim} 1.9\; TeV)$ and the squarks of the first two
families' $(1.2\; TeV \stackrel{_<}{_\sim} m_{\tilde{u},\tilde{d}}
\stackrel{_<}{_\sim} 2.1\; TeV)$. The discovery
of the top quark will narrow the allowed range and is therefore
eagerly awaited.

\vskip 1cm

{\noindent {\bf Acknowledgements}}:  We acknowledge useful
discussions with X.M. Wang and M. Drees. B.A. thanks Xerxes Tata
for a conversation.  We gratefully acknowledge the use of the
package JACOBI from Ref. \cite{numrec}
\pagebreak

\pagebreak

\noindent
{\large\bf Figure Captions}
\vspace{.4in}

\begin{tabular}{ll}

{\bf Fig. 1}. &  Contours of $M_{\frac{1}{2}}^{min}|_{m_A}$
and $m_{0}^{min}|_{m_A}$ for $m_A=M_Z$.\\
\\
{\bf Fig. 2}. & Plot of bino purity as a function of $m_t(m_t)$.\\
\\
{\bf Fig. 3}. &  Plot of $m_{h^\circ}$ as a function of $m_A$.\\
\\
{\bf Fig. 4}. &  Plot of $m_{h^\circ}^{max}$ as a function
of $m_t(m_t)$.\\
\\
{\bf Fig. 5}. &  Plot showing maximum allowed values of the higgs
masses,\\
& $m_{A}^{max}$, $m_{H^\circ}^{max}$ and $m_{H^{\pm}}^{max}$.
\end{tabular}

\begin{thebibliography}{abcdef}
\begin{small}
\bibitem{cdf} CDF Collaboration, Fermilab, Conf. - 93/212-E, 1993.

\bibitem{aleph} M. Felcini, CERN - PPE/92-208, Dec. 1992.

\bibitem{lang} For a recent review, see P. Langacker,
``Precision tests of the standard model,'' University of
Pennsylvania preprint, UPR-0555T, 1993; See also J. Ellis, G.L. Fogli
and E. Lisi, Nucl. Phys. B393 (1993) 3.

\bibitem{hn} For reviews see, H. P. Nilles, Phys. Rep. 110 (1984) 1;
H. E. Haber and G. L. Kane, Phys. Rep. 117 (1985) 75;
F. Zwirner, Lectures delivered at the Trieste Summer School in
High Energy Physics and Cosmology, 1991.

\bibitem{als} B. Ananthanarayan, G. Lazarides and Q. Shafi,
Phys. Rev. D44 (1991) 1613;  Q. Shafi and B. Ananthanarayan,
``Will LEPII narrowly miss the Weinberg-Salam-Higgs Boson?''
{\it 1991 Summer School in High Energy Physics and Cosmology}, pp.
233, E Gava, {\it et al.} eds., World Scientific, Singapore, 1992;
ibid, Paschos Symposium Proc. (1991).

\bibitem{als2} B. Ananthanarayan, G. Lazarides and Q. Shafi,
Phys. Lett. 300B (1993) 245.

\bibitem{il}  K. Inoue, A. Kakuto, H. Komatsu and S. Takeshita, Prog.
Theo. Phys., 68 (1982) 927; K. Inoue, A. Kakuto, H. Komatsu and S. Takeshita,
Prog. Theo. Phys., 71 (1984) 413;
L. E. Ib\'{a}\~{n}ez and C. L\'{o}pez, Nucl. Phys. B233(1984) 511;
L. E. Ib\'{a}\~{n}ez, C. L\'{o}pez and C. Mu\~{n}oz,
Nucl. Phys. B256(1985) 218;
M. Drees and M. M. Nojiri, Nucl. Phys. B369 (1992) 54 and references
therein.

\bibitem{hrs} L.J. Hall, R. Rattazzi and U. Sarid, LBL preprint 33997
(1993); V. Barger et al, Univ. of Madison preprint MAD/PH/781 (1993);
P. Langacker and N. Polonsky, Univ. of Pennsylvania preprint
UPR-0556-T (1993); H. Arason et al., Phys. Rev. Lett., 67 (1991)
2933; M. Carena et al., Nucl. Phys. B406 (1993) 59; S. Kelley et al.,
Phys. Lett. B278 (1992) 140.

\bibitem{ss} Q. Shafi and R.K. Schaefer,
Nature 359 (1992) 199 and references therein.

\bibitem{mk} K. Olive and M. Srednicki, Phys. Lett. 230B (1989) 78;
K. Griest, M. Kamionkowski and M. S. Turner,
Phys. Rev. D41 (1990), 3565;
M. Drees and M. M. Nojiri, Phys. Rev. D47 (1993) 376; J. McDonald, K.
Olive and M. Srednicki, Phys. Lett. B283 (1992) 80.

\bibitem{grz} G. Gamberini, G. Ridolfi and F. Zwirner,
Nucl. Phys. B331 (1990) 331.

\bibitem{hk} H. Komatsu, Phys. Lett. 215B (1988) 323.

\bibitem{gl} J. Gasser and H. Leutwyler, Phys. Rep. 87 (1992) 77.

\bibitem{higgs} J. Ellis, G. Ridolfi and F. Zwirner,
Phys. Lett. 257B (1991) 83; J. Ellis, G. Ridolfi and F. Zwirner,
Phys. Lett. 262B (1991) 477; Y. Okada, M. Yamaguchi and
T. Yanagida, Prog. Theor. Phys. Lett 85 (1991) 1;
Y. Okada, M. Yamaguchi and T. Yanagida, Phys. Lett. 262B (1991) 54;
R. Barbieri, M. Frigeni and F. Caravaglios, Phys. Lett. 258B (1991)
167; J. R. Espinosa and M. Quir\'{o}s, Phys. Lett. 266B (1991) 389;
H. E. Haber and R. Hempfling, Phys. Rev. Lett. 66 (1991) 1815;
P. N. Pandita, ``Radiative corrections to Higgs Boson Masses in
Supersymmetric Models,'' Talk given at the X DAE High Energy Symposium
and references therein.

\bibitem{cleo} CLEO Collaboration, Washington APS Meeting, April
1993.

\bibitem{hewett} J.L. Hewett, Phys. Rev. Lett. 70 (1993) 1045; V.
Barger, M.S. Berger and R.J.N. Phillips, Phys. Rev. Lett. 70 (1993)
1368.

\bibitem{numrec} {\it Numerical Recipes},  W. H. Press,
B. P. Flannery, S. A. Teukolsky and W. T. Vetterling,
Cambridge University Press, Cambridge, 1986.

\end{small}
\end{thebibliography}
\end{document}